\documentclass[a4paper,11pt]{article}
\usepackage{graphicx}
\usepackage{amsfonts}
\usepackage{amsmath}
\usepackage{enumerate}
\usepackage{subfigure}
\usepackage{multirow}
\usepackage{psfrag}
\usepackage{tikz}

\newcommand{\bx}{\boldsymbol{x}}

\newcommand{\by}{\boldsymbol{y}}
\newcommand{\bz}{\boldsymbol{z}}
\newcommand{\btheta}{\boldsymbol{\theta}}
\newcommand{\bbeta}{\boldsymbol{\beta}}

\newcommand{\bmu}{\boldsymbol{\mu}}
\newcommand{\bSigma}{\boldsymbol{\Sigma}}
\newcommand{\bgamma}{\boldsymbol{\gamma}}

\newcommand{\nr}{n \rightarrow \infty}

\begin{document}
\title{Model comparison with missing data using MCMC and importance sampling}
\author{Panayiota Touloupou$^1$ (University of Warwick), Naif Alzahrani$^2$ (Lancaster University), \\ Peter Neal$^2$ (Lancaster University),  Simon E.F. Spencer$^1$ (University of Warwick) \\ and Trevelyan J. McKinley$^3$ (University of Exeter)}

\maketitle

\footnote{1: Department of Statistics, University of Warwick, Coventry, CV4 7AL, UK; \\ 2: Department of Mathematics and Statistics,
Lancaster University, Lancaster, LA1 4YF, UK; \\ 3: College of Engineering, Mathematics and Physical Sciences, University of Exeter, Penryn Campus, Penryn, Cornwall, TR10 9EZ, UK}

\begin{abstract}
Selecting between competing statistical models is a challenging problem especially when the competing models are non-nested. In this paper we offer a simple solution by devising an algorithm which combines MCMC and importance sampling to obtain computationally efficient estimates of the marginal likelihood which can then be used to compare the models. The algorithm is successfully applied to longitudinal epidemic and time series data sets and shown to outperform existing methods for computing the marginal likelihood.

\end{abstract}


\section{Introduction} \label{S:intro}

The central pillar of Bayesian statistics is Bayes' Theorem. That is, given a parameteric model $\mathcal{M}$ with parameters $\btheta = (\theta_1, \ldots, \theta_d)$ and data $\mathbf{x} = (x_1,x_2,\ldots, x_n)$, the joint distribution of $(\btheta, \mathbf{x})$ satisfies
\begin{eqnarray} \label{eq:Bayes}
 \pi (\btheta | \mathbf{x}) \pi (\mathbf{x}) = \pi (\mathbf{x} | \btheta) \pi (\btheta).
\end{eqnarray}
The four terms in \eqref{eq:Bayes} are the posterior distribution ($\pi (\btheta | \mathbf{x})$), the marginal likelihood ($\pi (\mathbf{x})$), the likelihood ($\pi (\mathbf{x} | \btheta)$) and the prior distribution ($\pi (\btheta)$). The terms on the right hand side of \eqref{eq:Bayes} are usually easier to derive than those on the left hand side. The statistician has considerable control over the prior distribution and this can be chosen pragmatically to reflect prior beliefs and to be mathematically tractable. For many statistical problems the likelihood can easily be derived. However, the quantity of primary interest is usually the posterior distribution. Rearranging \eqref{eq:Bayes} it is straightforward to obtain an expression for  $\pi (\btheta | \mathbf{x})$ so long as the marginal likelihood can be computed. This involves computing
\begin{eqnarray} \label{eq:marge} \pi (\mathbf{x}) = \int \pi (\mathbf{x} | \btheta) \pi (\btheta) d \btheta \end{eqnarray} which is only possible for a relatively small set of simple models.

A key solution to being unable to obtain an analytical expression for the posterior distribution is to obtain samples from the posterior distribution using Markov chain Monte Carlo (MCMC), \cite{Metro}, \cite{Hastings}. A major strength of MCMC is that it circumvents the need to compute  $\pi (\mathbf{x})$ and this has led to its widespread use in Bayesian statistics over the last 25 years. However, for model comparison and other purposes we might be interested in $\pi (\mathbf{x})$. For example, the computation of Bayes Factors, \cite{KR95a}, require the marginal likelihoods for the competing models. In \cite{Chib} a simple rewriting of \eqref{eq:Bayes} was exploited to obtain estimates of the marginal likelihood using output from a Gibbs sampler. This has been extended in \cite{CJ01} and \cite{Chen} to general Metropolis-Hastings algorithm. It should be noted that model comparison using reversible jump (RJ)MCMC, \cite{Green} is possible for a range of models and again avoids the computation of the marginal likelihood. RJMCMC works well for nested models where it is straightforward to define a good transition rule for models with different parameters. Using RJMCMC becomes harder when data augmentation is required within the MCMC algorithm so that the dimension of the parameters (and augmented data) becomes large, especially where the data augmentation required for the different models have a different structure. An example of this is the comparison between INAR models \cite{NSR} and Poisson regression models \cite{DDS} for low count time series data in Section \ref{S:TS}. Our failure to devise a suitable reversible jump schema between these two models provided the motivation for the current work.

The remainder of the paper is structured as follows. In Section \ref{S:gen} we motivate and outline the estimation procedure devised in this paper. The estimation procedure involves three stages: estimation of $ \pi (\btheta | \mathbf{x})$ using MCMC; approximating $\pi (\btheta | \mathbf{x})$ by $q (\btheta)$, the probability density function of a parametric distribution; and finally the estimation of $\pi (\mathbf{x})$ using importance samples from $q (\btheta)$. We note that the three stages can be incorporated into a single automatic algorithm. We also discuss briefly the properties of the estimator of $\pi (\mathbf{x})$. In Section \ref{S:epi} we apply the methodology to an epidemic example, for the transmission of {\it Streptococcus pneumoniae} \cite{Melegaro}. We compare the performance of the algorithm developed in Section \ref{S:gen} with three alternative methods for computing the marginal likelihood; Chib's method \cite{Chib}, power posteriors method \cite{Friel} and the harmonic mean \cite{Newton}. The new algorithm is shown to significantly outperform the alternative methods and to produce tighter Bayes Factor estimates than those obtained using RJMCMC for comparing competing models. In Section \ref{S:TS} we use the algorithm to choose between competing models where as noted above, RJMCMC is no longer a practical solution. Finally in Section \ref{S:conc} we briefly discuss extensions and limitations of the algorithm.



\section{Algorithm for computing the marginal likelihood} \label{S:gen}

In this Section we provide the motivation and generic framework for
computing the marginal likelihood (model evidence). The details of how this is implemented are problem
specific and are thus postponed to Sections \ref{S:epi} and \ref{S:TS}.

The first observation is that we can rewrite \eqref{eq:marge} as
\begin{eqnarray} \label{eq:marg:3}
\pi (\mathbf{x}) = \int_{\btheta} \pi (\mathbf{x} | \btheta)
\frac{\pi (\btheta)}{q (\btheta)} q (\btheta) \, d \btheta,
\end{eqnarray}
where $q (\btheta)$ denotes a $d$-dimensional probability density
function. We assume that if $\pi (\btheta)>0$ then $q (\btheta)
>0$. Then an unbiased estimator, $\hat{P}_q$ of $\pi
(\mathbf{x})$ is obtained by sampling $\btheta_1, \btheta_2, \ldots,
\btheta_N$ from $q (\btheta)$ and setting
\begin{eqnarray} \label{eq:marg:4}
\hat{P}_q = \frac{1}{N} \sum_{i=1}^N \pi (\mathbf{x} | \btheta_i)
\frac{\pi (\btheta_i)}{q (\btheta_i)}.
\end{eqnarray}
Thus $\hat{P}_q$ is an importance sampled estimate of $\pi (\mathbf{x})$ and
the effectiveness of the estimator given by \eqref{eq:marg:4}
depends upon the variability of $\pi (\mathbf{x} | \btheta_1) \pi
(\btheta_1)/q (\btheta_1)$. The optimal choice of $q (\cdot)$ is $q
(\btheta) = \pi (\btheta | \mathbf{x})$ since then for any $N \geq
1$, by \eqref{eq:Bayes},
\begin{eqnarray} \label{eq:marg:5}
\hat{P}_q &=& \frac{1}{N} \sum_{i=1}^N \pi (\mathbf{x} | \btheta_i)
\frac{\pi (\btheta_i)}{\pi (\btheta_i |\mathbf{x})} \nonumber \\
&=& \frac{1}{N} \sum_{i=1}^N \frac{\pi ( \btheta_i |\mathbf{x})
\pi (\mathbf{x})}{\pi (\btheta_i |\mathbf{x})} \nonumber \\
&=& \frac{1}{N} \sum_{i=1}^N \pi (\mathbf{x}) = \pi (\mathbf{x}).
\end{eqnarray}
However, if we know $\pi (\btheta | \mathbf{x})$ then there is no
need to estimate $\pi (\mathbf{x})$ but the above is useful, in
that, it tells us that we should choose  $q (\btheta)$ to
approximate $\pi (\btheta | \mathbf{x})$ as closely as possible.

The above calculations raise the important question
of how do we get a reasonable approximation to $\pi (\btheta |
\mathbf{x})$? The solution proposed in this paper is to use MCMC to
obtain samples from $\pi (\btheta | \mathbf{x})$ and then to use the
samples to construct an approximating sampling distribution $q
(\btheta)$. For most statistical models the likelihood times the prior is unimodal for sufficiently large $n$. In these circumstances for large $n$, the posterior distribution of $\btheta$ is almost always approximately Gaussian with mean $\hat{\btheta}$, the posterior mode, and covariance matrix $\Sigma = - \mathcal{I} (\hat{\btheta})^{-1}$, where $\mathcal{I} (\btheta)$ denotes the Fisher information evaluated at $\btheta$. That is, we have a central limit theorem type behaviour similar to that observed for maximum likelihood estimators as $\nr$. This central limit theorem approximation is implicitly behind the Laplace approximations of integrals used in \cite{TK86a}, (2.2) and \cite{GD94a}, (8).
Thus where the posterior distribution is uni-modal a multivariate Gaussian or $t$-distribution based on the sample using the sample mean and covariance matrix is a natural choice.
Let $\omega (\btheta) = \pi (\btheta |\mathbf{x})/ q(\btheta)$, then $\hat{P}_q = \frac{1}{N} \pi (\mathbf{x}) \sum_{i=1}^n \omega (\btheta_i)$. Thus
\begin{eqnarray}
\label{eq:marg:var1}
var_q (\hat{P}_q)& =& \frac{1}{N} \pi (\mathbf{x})^2 var_q (\omega (\btheta_1)).\end{eqnarray}
Thus we require $var_q (\omega (\btheta_1))$ as small as possible and in particular that $\sup_{\btheta} \omega (\btheta) < \infty$. This latter condition is satisfied if $\sup_{\btheta} \pi (\mathbf{x} | \btheta) < \infty$ and there exists $\epsilon >0$ such that for all $\btheta$, $q
(\btheta) \geq \epsilon  \pi(\btheta)$. Note that a mixture distribution combining an empirical approximate distribution with the prior can be used to circumvent this problem. In Section \ref{S:epi}, we compare alternative choices of $q (\cdot)$ and show that all the above suggestions perform well, especially the mixture distribution proposal.

The above discussion assumes that $\pi (\mathbf{x} | \btheta)$ is analytically available. This will not always be the case. However, the estimator $\hat{P}_q$ in \eqref{eq:marg:4} remains unbiased if  $\pi (\mathbf{x} | \btheta_i)$ is replaced by an unbiased estimator $\widehat{\pi (\mathbf{x} | \btheta_i)}$. The variance of $\hat{P}_q$ will increase to take into account the uncertainty in the estimation of $\pi (\mathbf{x} | \btheta_i)$. In the estimation of $\pi (\mathbf{x} | \btheta_i)$ it will often be helpful to use data augmentation with augmented data $\mathbf{y}$, and indeed samples from $\pi (\btheta | \mathbf{x})$ will often be obtained as a marginal of $\pi (\btheta, \mathbf{y} | \mathbf{x})$ generated by an MCMC algorithm. The choice of data augmentation and the resulting estimation of $\pi (\mathbf{x} | \btheta_i)$ will be problem specific. For example, in Section \ref{S:TS} a particle filter is used to estimate $\pi (\mathbf{x} | \btheta_i)$ for the Poisson regression model, whereas for the INAR model $\pi (\mathbf{x} | \btheta_i)$ is readily available, despite the need to use data augmentation for an efficient MCMC algorithm. Note that $\mathbf{y}$ can be high dimensional and seeking a good approximation for  $\pi (\btheta, \mathbf{y} | \mathbf{x})$ is rarely possible. Therefore we focus on approximating $\pi (\btheta | \mathbf{x})$ and then sampling $\mathbf{y}$ given $\btheta$ and $\mathbf{x}$ in a systematic and appropriate manner as required.

The algorithm we implement can be summarised as follows with the details of implementation being problem specific.
\begin{enumerate}
\item Obtain samples from $\pi (\btheta | \mathbf{x})$ using an MCMC sampler.
\item Construct a proposal distribution $q (\btheta)$ for the importance sampler using the samples from $\pi (\btheta | \mathbf{x})$ as guidance.
\item Compute the importance sampling estimate $\hat{P}_q$, using an unbiased estimator for $\pi (\mathbf{x}| \btheta)$ if necessary.
\end{enumerate}

We briefly comment on the statistical and computational efficiency of the above algorithm. Stage 1 is simply an MCMC sampler and this needs to be run for sufficient iterations to obtain a reasonable sample from the posterior distribution. Thus the computational cost is that associated with running the MCMC sampler. Stage 2 will generally be computationally cheap as it involves obtaining an approximate distribution for $\pi (\btheta | \mathbf{x})$.
We can encounter typical issues for importance sampling as $d$, the dimension of $\btheta$, increases it is harder to obtain a good approximation $q (\cdot)$ to the posterior distribution, although for $d \leq 10$ the $t$-distribution approximation appears to be more than adequate. Finally, in stage 3 we require $N$ to be sufficiently large that $var_q (\hat{P}_q)$ is small. This will depend upon how good an approximation $q (\cdot)$ is for the posterior density and the variability in the estimate of $\pi (\mathbf{x}| \btheta)$. For the examples studied in Sections \ref{S:epi} and \ref{S:TS}, $N$ between 10000 and 25000 was more than adequate. The computational costs of stages 1 and 3 are often comparable but will depend upon the effort required to estimate $\pi (\mathbf{x}| \btheta)$ efficiently.


%

%
%

\section{Epidemic model} \label{S:epi}

\subsection{Introduction} \label{ss:S:intro}

Epidemiological data from infectious disease studies are very often gathered longitudinally, where a collection of individuals are sampled through time. Inferences for this type of data are complicated by the fact that the data are usually incomplete, in the sense that the times of acquiring and clearing infection are not directly observed, making the evaluation of the model likelihood intractable. One solution to this problem is to use data augmentation methods, in which the unobserved infection and recovery times are imputed from the observed data at the cost of considerable extra computational effort. In the Bayesian framework this is facilitated by the use of Markov chain Monte Carlo (MCMC) algorithms, which enable the data imputation and parameter estimation to be performed simultaneously \cite{O'Neill,Streftaris}. The availability of efficient MCMC algorithms has led to them becoming the most prevalent techniques for analysing data on partially observed epidemics.

In \cite{NR04a} and \cite{ONM05} model selection for the modes of disease transmission have been studied using reversible jump MCMC, whilst in \cite{KON14a} Bayes factors are computed using path sampling-based algorithms to compare competing models.  We consider the problem of comparing a set of candidate models using the approach outlined in Section \ref{S:gen} for estimating the marginal likelihood of the competing models. The main complication is the incompleteness of the data making computation of the likelihood intractable and the need therefore to find effective way of estimating the likelihood.

The rest of this section is structured as follows. The longitudinal pneumococcal carriage study motivating this analysis is described in Section \ref{ss:S:Pneumo}. We then introduce the notation and the model structure in Sections \ref{ss:S:MCMC algo} and \ref{ss:S:mlcalculation}. Simulated data are provided in Section \ref{ss:S:sim} to illustrate the implementation, performance and applicability of the proposed method. In Section \ref{ss:S:comp} we apply the model selection procedures to  simulated data sets (where the true model is known), in order to demonstrate how the model selection can be used to quantify evidence in favour of competing epidemiological hypotheses. Throughout we demonstrate the effectiveness of the proposed approach against a range of alternatives.

\subsection{Pneumococcal carriage study and transmission model} \label{ss:S:Pneumo}

A longitudinal household study of preschool children under 3 years old and all household members was conducted in the United Kingdom from October 2001 to July 2002 \cite{Hussain}. The size of the families varied from 2 to 7, although in the most there were 3 or 4 members. All family members were examined for {\it Streptococcus pneumoniae} carriage (Pnc) using nasopharyngeal swabs once every 4 weeks over a 10-month period. The carriage status of each individual was recorded at each occasion as 1, if a carrier, 0 if a non-carrier and 9 when either the swab was not taken or the laboratory result was not reported.

Following \cite{Melegaro}, we consider a model for transmission of Pnc within a household. At any given time, an individual is assumed to be in either the non-carrier, susceptible state 0, or the infectious carrier state 1. The population is divided into two age groups, children under 5 years old and everyone else greater than 5 years (whom for brevity we refer to as `adults'), denoted by $i=1, 2$, respectively. Let $I_1(t)$ and $I_2(t)$ denote the numbers of carrier children and carrier adults in the household at time $t$. The transition between state 0 and 1 is referred to as an infection and the reverse transition is referred to as clearance. The transition probabilities between states in a short time interval $\delta t$ are defined for an individual in the age group $i$:
  \begin{align} \label{eq:Pneumo:1}
P(\text{Infection in $(t,t+\delta t]$}) &=1-\exp\left\{-\left( k_i + \frac{\beta_{1i}\, \text{I}_1(t)+ \beta_{2i} \, \text{I}_2(t)}{ (z-1)^w} \right)\cdot \delta t\right\} \\
\label{eq:Pneumo:2}
P(\text{Clearance in $(t,t+\delta t]$}) &=1-\exp(-\mu_i \cdot \delta t),   \end{align}
where $\mu_i$ and $k_i$ are the clearance and the community acquisition rates respectively for age group $i$ and $z$ is the household size. The rate $\beta_{ij}$ is the transmission rate from an infected individual in age group $i$ to an uninfected individual in age group $j$. The term $(z-1)^w$ in (\ref{eq:Pneumo:1}) represents a density correction factor, where $w$ corresponds to the level of density dependence and $(z-1)$ is the number of other family members in a household size $z$. For example, $w=1$ represents frequency dependent transmission, where the average number of contacts is equal for each individual in the population. Finally, the probability of infection at the initial swab is assumed to be $\pi_1$ for children under 5 years old and $\pi_2$ for individuals 5 years and older. We refer to this model as $\mathcal{M}_1$.

These definitions allow the carriage within a household to be viewed as a discrete time Markov chain, with time step $\delta t$, where the carriage status of each individual depends only on the carriage status of all household members at the previous time point. Because of the dependency between individuals in the same household, a state in the Markov chain consists of the binary vector of states of all of the individuals in the household. The presence of unobserved events, that may have occurred in between swabbing intervals, has been discussed previously \cite{Auranen}, and must be considered in setting up the model. The approach adopted in this article to overcome this issue is to use Bayesian data augmentation methods. Model fitting is performed within a Bayesian framework using an MCMC algorithm, imputing the unobserved carriage states of each household.

Let $O_j\subseteq\{1,2,\dots,T\}$ denote the set of prescheduled observations times of household $j = 1, 2, \ldots, J$ and let $U_j=\{1,2,\dots,T\}\setminus O_j$ denote the unobserved times. Let $\bx_{j,t}$ be the binary vector of carriage statuses for individuals in household $j$ at observation time $t$. The observed longitudinal data $\mathbf{X}=[\bx_{j,t}]_{t\in O_j;j=1,\dots,J}$ consists of the household carriage statuses $\bx_{j,t}$ at the observations times. Similarly let $\by_{j,t}$ be the corresponding latent carriage status of household $j$ at time $t\in U_j$, and form the corresponding missing data matrix $\mathbf{Y}=[\by_{j,t}]_{t\in U_j;j=1,\dots,J}$. Let $\boldsymbol{\theta}$ denote the vector of model parameters, including the rates of acquiring and clearing carriage, the density correction $w$ and the initial probabilities of carriage.

\subsection{Markov chain Monte Carlo algorithm}
\label{ss:S:MCMC algo}

In the Bayesian approach, the missing data is represented as a nuisance parameter and inferred from the observed data like any other parameter. The joint posterior density of the latent carriage statuses $\mathbf{y}$, and the model parameters $\boldsymbol{\theta}$ can be factorized as:
\begin{align*}
\pi(\mathbf{y},\boldsymbol{\theta} \mid  \mathbf{x}) &\propto P( \mathbf{y}, \mathbf{x}\mid\boldsymbol{\theta}) \,  \pi(\boldsymbol{\theta})  \\
&=\pi(\btheta)\prod_{j=1}^J\prod_{t=1}^T P(\bz_{j,t}|\bz_{j,t-1},\btheta),
\end{align*}
where $\bz_{j,t}$ equals $\bx_{j,t}$ if $t\in O_j$; $\by_{j,t}$ if $t\in U_j$ and $\emptyset$ if $t=0$. This factorization is based on the assumption that conditionally on the model parameters, the carriage process is assumed to be independent across households.

In order to simulate from the posterior distribution, we construct an MCMC algorithm that employs both Gibbs and Metropolis-Hastings updates. The main emphasis is on sampling the unobserved carriage process $\mathbf{y}$, which we do using a Gibbs step via the Forward
Filtering Backward Sampling (FFBS) algorithm \cite{Carter}. In the first part of this algorithm, recursive filtering equations \cite{Anderson} are used to calculate $P(\by_{j,t}\mid \bz_{j,t+1},\bx_{j,O_j\cap\{1:t\}},\btheta)$ for each $t\in U_j$ working forwards in time. The second part then works backwards through time, simulating $\by_{j,t}$ from these conditionals, starting with $t=\max(U_j)$ and ending with $t=\min(U_j)$. The model parameters $\pi_1$ and $\pi_2$ are updated using Gibbs updates and the remaining parameters are updated jointly using an adaptive Metropolis-Hastings random walk proposal \cite{RobertsAdaptiveMCMC}.

\subsection{Marginal likelihood estimation via importance sampling}
\label{ss:S:mlcalculation}

The availability of the full conditional distribution of the missing data $P(\mathbf{y} | \mathbf{x}, \boldsymbol{\theta})$ from the FFBS algorithm allows the missing data component $\mathbf{y}$ to be updated using a Gibb's step in the MCMC algorithm. This full conditional can be exploited further in the estimation of the marginal likelihood. In step 3 of the algorithm described in Section \ref{S:gen}, we require $P(\mathbf{x}|\btheta)$ to form the importance sampling estimator. Using Bayes' Theorem we can rewrite this as
\begin{align}
\label{eq:pXtheta}
P( \mathbf{x}|\boldsymbol{\theta}) & = \frac{P(\mathbf{x} | \mathbf{y}, \boldsymbol{\theta}) \,  P(\mathbf{y} | \btheta) } {P(\mathbf{y} | \mathbf{x}, \boldsymbol{\theta})}  = \frac{P( \mathbf{x}, \mathbf{y} | \boldsymbol{\theta}) } {P(\mathbf{y} | \mathbf{x}, \boldsymbol{\theta})}.
\end{align}
Therefore evaluation of $P( \mathbf{x}|\boldsymbol{\theta})$ at the point $\btheta$ can be done by sampling from $\mathbf{y}|(\mathbf{x},\btheta$) and evaluating the right-hand-side of (\ref{eq:pXtheta}).

Our approach proceeds as follows. In step 1 we use MCMC to obtain samples from the joint posterior of $\btheta$ and $\mathbf{y}$. In step 2 we fit a multivariate normal distribution to the posterior samples for $\btheta$ only, and use it to construct a normalised proposal density $q(\btheta)$. In step 3, we obtain $N$ samples from $q(\btheta)$ and for each sample $\btheta_i$ we obtain a corresponding sample for the missing data $\mathbf{y}_i$ using the Forward Filtering Backward Sampling algorithm. We then use these samples to calculate the importance sampling estimator of the marginal likelihood given below.
\begin{align}
\label{eq:ISesti}
\widehat{P}_q(\mathbf{x})
&=\frac{1}{N}\sum_{i=1}^N \frac{P( \mathbf{x}, \mathbf{y}_i | \btheta_i) } {P(\mathbf{y}_i | \mathbf{x}, \btheta_i)} \frac{\pi(\btheta_i)}{q(\btheta_i)}.
\end{align}

The choice of $q (\btheta)$ is important for the accuracy and computational efficiency of the importance sampling approach.
As discussed earlier, we want $q(\btheta)$ to be a good approximation of $\pi(\btheta|\mathbf{x})$ but with heavier tails to ensure that the variance of $\hat{P}_q$ is small. We therefore investigate a range of proposals distributions based on a fitted multivariate normal distribution with mean $\bmu$ and covariance matrix $\bSigma$ based on the MCMC output. These include drawing $\btheta$ from $IS_{N_j}: N (\bmu, j \bSigma)$ $(j=1,2,3,4)$ (multivariate Normal distribution with different variances); $IS_{\rm{mix}}: q(\btheta)=0.95 \times N (\btheta;\bmu,  \bSigma) + 0.05\times\pi (\btheta)$ (mixture of a multivariate Normal density and the prior); and $IS_{t_d}: t_d (\bmu, \bSigma)$ $(d=4,6,8,10)$ (multivariate Student's $t$ distribution with $d$ degrees of freedom, mean $\bmu$ and covariance matrix $\frac{d}{d-2}\bSigma$ (if $d>2$)).

We also performed a comparison between the importance sampling approach and a range of commonly used alternative approaches to marginal likelihood estimation. These approaches include the harmonic mean \cite{Newton}, Chib's method \cite{Chib,CJ01} and the power posteriors method (also referred to as path sampling), \cite{Friel}. Details of the computation of these estimators are given in the appendix.

\subsection{Simulation study}\label{ss:S:sim}

We consider the problem of estimating the marginal likelihood under the model introduced in Section \ref{ss:S:Pneumo}, using the methods described above. These estimators were evaluated on synthetic data analogous to the real data  in \cite{Melegaro}. More specifically, the parameter values were based on the maximum likelihood estimates from the analysis of Pnc data;  parameters were chosen to be $k_1 = 0.012, \, k_2 = 0.004,\,  \beta_{11} = 0.047,\,   \beta_{12} = 0.106,\,  \beta_{21} = 0.005,\,  \beta_{22} = 0.048, \,  \mu_1 = 0.020, \, \mu_2 = 0.053,\,  w=1.184, \, \pi_1 = 0.425$ and $ \pi_2 = 0.095$. We set the time-interval $\delta t = 7$. Only complete family transitions, where the infection state of all household members was known on two consecutive observations, were used previously \cite{Melegaro} (51$\%$ of the full dataset). Although our approach could easily handle the missing data, for comparability we match the number of complete transitions by family size and number of adults to generate our data set; a total of 66 families comprising 260 individuals including 94 children under 5 years. The simulations were designed so that real and simulated datasets have the same sampling times. The hidden variable $\mathbf{y}$ consists of 1650 $\by_{j,t}$'s, comprising 6500 unobserved binary variables in total.

We compare the proposed importance sampling approach for estimating the marginal likelihood  (based on the 9 proposal densities) with Chib's method, the harmonic mean and the power posterior approaches. To compare the different methods on a fair basis, we chose
to dedicate equivalent amounts of computational effort for estimation of the log marginal likelihood, instead of fixing the total number of samples. Implementation details are given as follows.

The construction of the importance density was based on 25000 MCMC samples after a burn-in of 5000, obtained from the MCMC samplers described in Section \ref{ss:S:MCMC algo}. These posterior samples were used to estimate the reference parameters $\bmu$ and $\bSigma$ for a multivariate Student's $t$ or normal proposal density. The marginal likelihood estimate was then based on 25000 importance sampling draws from the obtained proposal density $q(\btheta)$, using the estimator in (\ref{eq:ISesti}). In order to apply the Chib's method, the same posterior sample was used for computing the high posterior density point. The log marginal approximation was produced by generating 22000 draws in each complete and reduced MCMC run, with the first 2000 draws removed as burn-in. Harmonic mean analysis was based on 50000 posterior samples, following a 3000 iteration burn-in. For the power posterior method, it was necessary to specify the temperature scheme and a pilot analysis (not counted in the computation cost) was used to choose 20 partitions on the unit interval. More details are given in the appendix.
The MCMC sampler was run for 2650 iterations for each temperature in the descending series, omitting the first 650 as burn-in, finishing with 2650 samples at $t=0$ (the prior). This results in approximately the same computational effort as the other methods.

Each procedure was repeated 50 times to provide an empirical Monte Carlo estimate of the variation in each approach. We also vary the total running time in order to investigate the effect of this on the accuracy of the marginal likelihood estimates, see Table 1 in the appendix. For each analysis method we used the same priors: Gamma(0.01,0.01) for the density factor $w$; Beta(1,1) for the initial probabilities of infection $\pi_1$ and $\pi_2$ and Gamma(1,1) for the remaining parameters.

 \begin{figure}[h!]
\centering
    \input{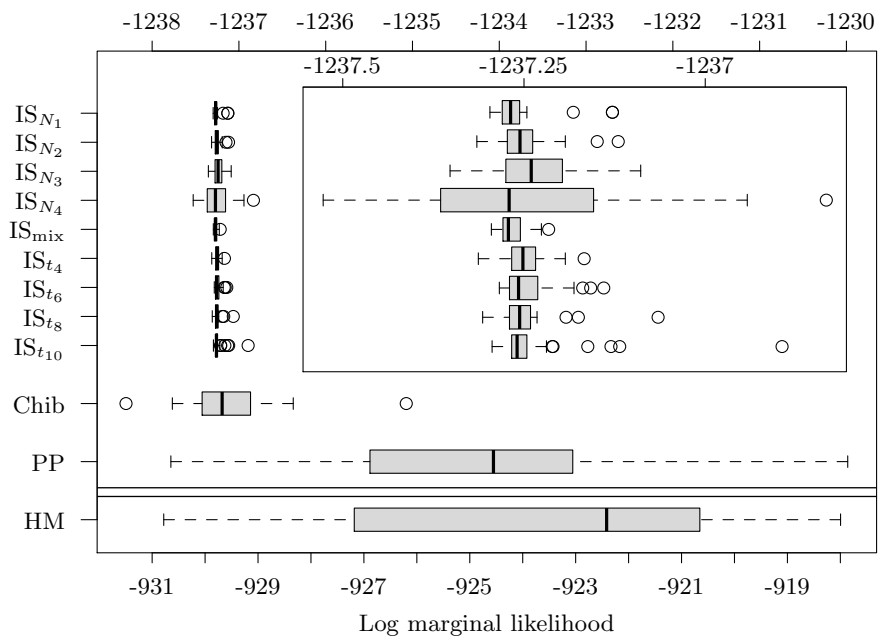}
    \caption{Variation of the log marginal likelihood estimates for model $\mathcal{M}_1$ over 50 replicates.}
 \label{fig:LogMargmain}
 \end{figure}

Figure \ref{fig:LogMargmain} indicates the variability of the twelve marginal likelihood estimators. Except for the harmonic mean, all the methods appear to have produced consistent estimates of the marginal likelihood. Chib's method produced better estimates of the marginal likelihood than the power posterior method, which is more computationally expensive than the other methods and therefore uses a small number of MCMC samples at each temperature, leading to large errors. However, the importance sampling approach described in Section \ref{S:gen} offers a great improvement in precision over the other competing methods. As can be seen in Figure \ref{fig:LogMargmain} the nine importance sampling approaches lead to nearly identical estimates for $\log \pi(\mathbf{x})$ with comparable standard errors, whereas the other three methods exhibit higher variability than these estimators. Moreover, increasing the number of MCMC samples, led to a decrease in the Monte Carlo standard errors of order $\mathcal{O}(\sqrt n)$, see Table 1 in the appendix, indicating that the variance of the corresponding estimators is finite.

The success of the importance sampling approach is not surprising since it explores the posterior distribution of parameters more efficiently than the other methods due to the independence of the samples drawn from the proposal density. On the basis of this example, the lowest variance estimator was obtained using the proposal density IS$_{\text{mix}}$ -- a mixture of the prior and the normal fitted to the posterior samples. Therefore, from now on we use this proposal density when estimating the log marginal likelihood via importance sampling.

\subsection{Model comparison} \label{ss:S:comp}

In this section, we apply the marginal likelihood estimation approaches to the problem of Bayesian model choice. We focus on their ability to distinguish between biologically motivated hypotheses concerning the dynamics of Pnc transmission. In particular we compare their performance against the established technique of Reversible Jump Markov Chain Monte Carlo (RJMCMC) and then demonstrate that the importance sampling approach can solve problems that are extremely challenging with RJMCMC. We show that using our approach it is possible to answer the epidemiological important question of how household size is related to transmission with extensive discussion given in the appendix.

Suppose that we wish to evaluate the evidence in favour of the community acquisition rates being equal for adults and children, in the hope of developing a more parsimonious model. We call the model described in Section \ref{ss:S:Pneumo}, in which children have community acquisition rate $k_1$ and adults have rate $k_2$, model $\mathcal{M}_1$. The nested model, in which $k_1=k_2$ is called $\mathcal{M}_2$. We generated realistic simulated datasets from each of these models and then used importance sampling, Chib's method, power posteriors, the harmonic mean and reversible jump MCMC to estimate the Bayes factor in favour of $\mathcal{M}_1$, denoted by $B_{12}$. As before, we used approximately the same computational effort for each of these approaches. For $\mathcal{M}_1$ we assumed $k_1 = 0.012$ and $k_2 = 0.004$, whilst for $\mathcal{M}_2$ we assumed $k_1 = k_2 = 0.008$.

Details of the RJMCMC algorithm for selecting between models $\mathcal{M}_1$ and $\mathcal{M}_2$ are given in the appendix. As before, the MCMC samples used for estimating Bayes factors with RJMCMC were designed to be comparable in computational effort with the other methods. Therefore, the RJMCMC chain was allowed a 30000 burn-in followed by 76000 samples. When the evidence is strongly in favour of one model, the RJMCMC will not move between models very often and can provide poor estimates of the Bayes factor. A variant of the method, called RJMCMC corrected (RJcor), can tackle this issue by assigning higher prior probability to the model that is visited less often. This probability is estimated as $\pi(\mathcal{M}_m) = 1 - \widehat{\pi}(\mathcal{M}_m \mid \mathbf{x})$, where $\hat{\pi}(\mathcal{M}_m \mid \mathbf{x})$ is obtained from a pilot run of RJMCMC with initial $\pi(\mathcal{M}_m) = 0.5$,  for $m=1,2$. For RJcor we did 30000 pilot iterations and then another 76000 iterations, of which 30000 were discarded as a burn in.

\begin{figure}[h!]
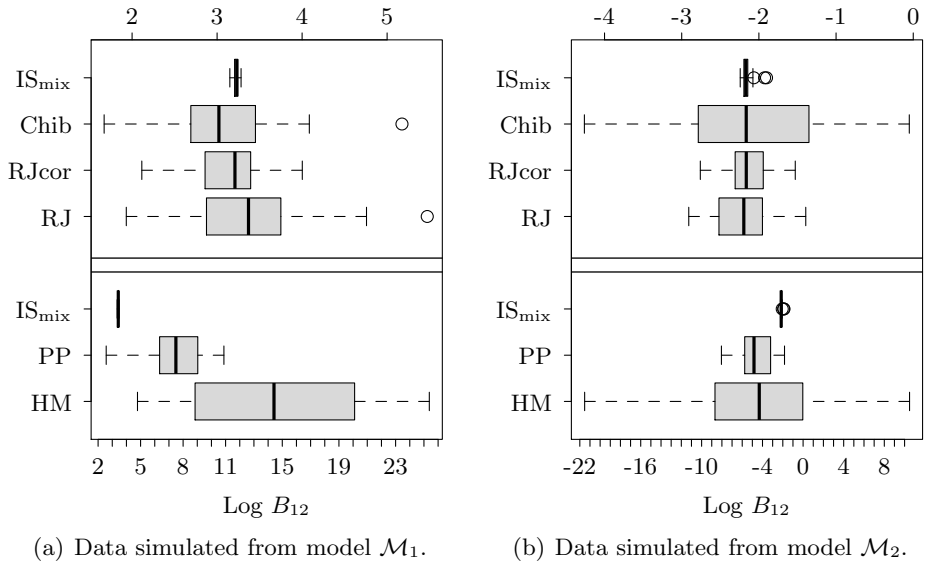

\centering
   \subfigure[Data simulated from model $\mathcal{M}_1$.]{ \input{Fig1secEpidModSela.tex} }
      \subfigure[Data simulated from model $\mathcal{M}_2$.]{ \input{Fig1secEpidModSelb.tex} }
    \caption{Variability of the log Bayes factor estimates based on 50 Monte Carlo repeats for the importance sampling method with mixture proposals (IS$_\text{mix}$), Chib's method, reversible jump MCMC (RJ), corrected reversible jump MCMC (RJcor), power posteriors (PP) and harmonic mean (HM) methods.}
    \label{fig:ExperimentA}
 \end{figure}

Figure \ref{fig:ExperimentA} provides a graphical representation of the variability in $\log(B_{12})$ over 50 repeats of each Monte Carlo approach. The plot highlights that the estimators based on  importance sampling were most accurate in both scenarios. The left panel of Figure \ref{fig:ExperimentA} gives results for data generated from $\mathcal{M}_1$.  Importance sampling, Chib and RJ methods lead to similar estimates, whereas power posterior and harmonic mean overestimated the log Bayes factor. Moreover, RJcor produced slightly more accurate estimates of the log Bayes factor than {\it vanilla} RJMCMC. All methods selected the correct model, with largest variation from the harmonic mean estimator. In the right panel of Figure \ref{fig:ExperimentA}, the results use data generated from model $\mathcal{M}_2$. Due to the huge variance in log($B_{12})$, the harmonic mean sometimes favoured the wrong model. Although the remaining methods correctly identified the true model, the importance sampling method again produced the most precise estimate of the Bayes factor.

\begin{figure}[h!]
\centering
  \input{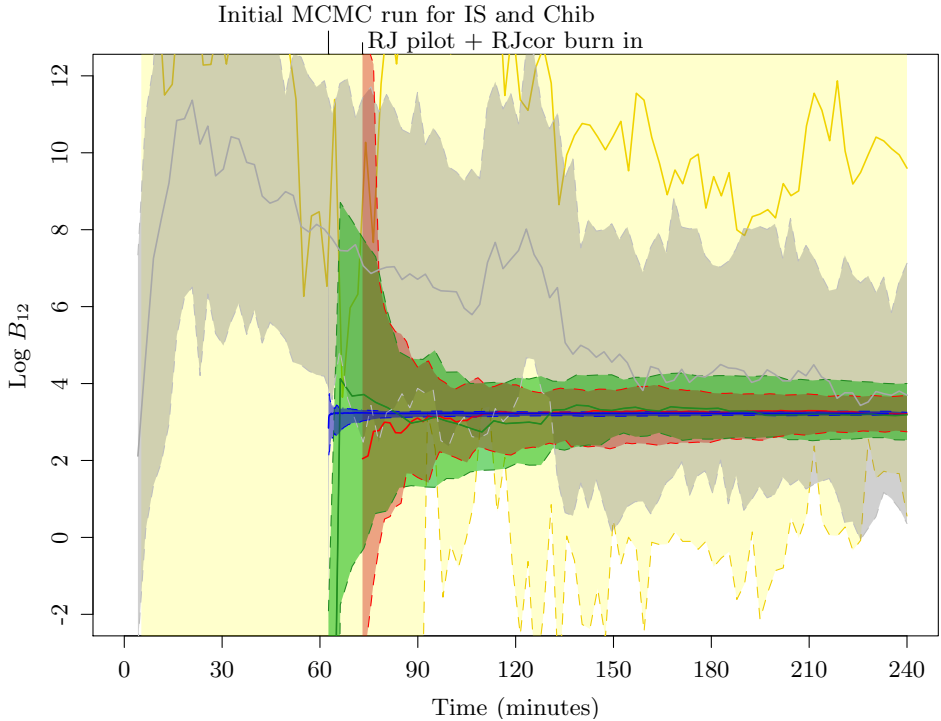}
    \caption{Evolution of log Bayes factor estimates in favour of model $\mathcal{M}_1$ as a function of computation time. The solid lines corresponds to the median and the shaded areas give the 95$\%$ credible intervals, estimated from 50 Monte Carlo replicates. Yellow represents the harmonic mean method, grey is for the power posterior, red and green correspond to RJMCMC corrected and Chib's methods respectively and blue represents the importance sampling approach with the mixture proposals.}
 \label{fig:CompTime}
 \end{figure}

Figure \ref{fig:CompTime} demonstrates the evolution of log Bayes factor in favour of $\mathcal{M}_1$ as a function of computation time using data generated from $\mathcal{M}_1$. The importance sampling estimator (in blue) converges much more rapidly than the other estimators, showing very tight credible intervals. Chib's method (in green) and corrected RJMCMC (in red) appear to converge to the same value, but more slowly and have wider CIs. The power posterior method gradually approaches the consensus estimate, requiring significantly more samples to stabilize. The harmonic mean estimator was heavily unstable and also provided much wider credible intervals than the other methods.

In the appendix further model comparison questions are considered and the strength of the importance sampling technique for answering these questions is further demonstrated. In particular, we consider heterogeneity in household transmission rates, density-dependence in within-household transmission and the amount of missing data.

\section{Time Series models} \label{S:TS}

\subsection{Introduction} \label{ss:TS:intro}

A key motivation for this work was to devise effective model
comparison tools for integer valued time series models. Probably the
two most common models for integer valued time series are the
integer autoregressive (INAR) model (see, for example, \cite{McK},
\cite{NSR}, \cite{ENS}) and the Poisson regression model (see, for
example, \cite{Zeger}, \cite{DDS}). A natural question is which of
these two models is most appropriate for a given data set, possibly
with the view to prediction of future values of the time series. The
very different nature of the two models makes constructing a
reversible jump MCMC algorithm (\cite{Green}) to move between the
two models impractical, hence the desire to estimate directly the
marginal likelihood of each model.

The remainder of this section is
structured as follows. In Section \ref{ss:TS:inar}, we consider the
INAR model. This is the simpler of the two models to estimate $\pi
(\mathbf{x})$ since whilst data augmentation is useful in the MCMC
algorithm to obtain samples from $\pi (\btheta | \mathbf{x})$ (see
\cite{NSR}), $\pi (\mathbf{x}| \btheta)$ can easily be computed
enabling \eqref{eq:marg:5} to be used.
In Section
\ref{ss:TS:pois} we turn to the far more challenging problem of
estimating $\pi (\mathbf{x})$ for the Poisson regression model.
Constructing an MCMC sampler in order to inform the choice of $q
(\btheta)$ is relatively straightforward. For sampling the augmented
data $\mathbf{y}$ a  particle filter is used. In Sections \ref{ss:TS:inar} and \ref{ss:TS:pois} we allow for covariates to be incorporated into the model. Finally, in Section \ref{ss:TS:comp} we apply the
methodology to two real life data sets and show that the preferred model differs between the data sets.

\subsection{INAR model} \label{ss:TS:inar}

An integer valued time-series $\{ X_t; - \infty < t < \infty \}$
is called an INAR$(p)$ process, if it satisfies the difference equation;
\begin{equation} \label{eq:inar} X_t = \sum_{i=1}^p \alpha_i \circ X_{t-i} + Z_t, \hspace{1.5cm} t \in \mathbb{Z},
\end{equation}
where $\alpha_i \circ$ are generalised Steutel and van Harn operators (see, \cite{SvH} and \cite{Latour97}) and $Z_t$ $(-
\infty < t < \infty)$ are independent and identically distributed
according to an arbitrary, but specified, non-negative integer
valued random variable $Z$. Often $\alpha_i \circ$ is taken to be a binomial operator (see, \cite{NSR}), where a binomial operator $\gamma \,
\circ$, for a non-negative integer-valued random variable, $W$
say, is defined as
\[ \gamma \circ W = \left\{ \begin{array}{ll} Bin (W , \gamma) \;
\; \; & W > 0, \\ 0 & W=0. \end{array} \right. \]
The most common choice for $Z_t$ is the Poisson distribution, $Z_t \sim {\rm Po} (\lambda)$.  In \cite{ENSb}, the INAR model is extended to allow $\alpha_i$ and $\lambda$ to be time dependent with $\alpha_i^t = \exp (\mathbf{z}_t^T \bbeta_i)/(1+\exp (\mathbf{z}_t^T \bbeta_i))$ and $\lambda^t = \exp (\mathbf{z}_t^T \bgamma)$, where $\mathbf{z}_t$ denote explanatory variables (covariates) at time-point $t$.

We assume that there are $n+p$ observations from the INAR$(p)$ process, labeled $x_{-(p-1)}, x_{-(p-2)}, \ldots, x_n$ with $\mathbf{x}_{IN} = (x_{-(p-1)}, x_{-(p-2)}, \ldots, x_0)$ denoting the initial $p$ observations and $\mathbf{x}= (x_1, x_2, \ldots, x_n)$ denoting the remaining $n$ observations. We work with $\pi ( \mathbf{x} | \btheta, \mathbf{x}_{IN})$ and compute the marginal likelihood  $\pi ( \mathbf{x} | \mathbf{x}_{IN})$ of the data $\mathbf{x}$ conditional upon the initial values $\mathbf{x}_{IN}$.

Computation of the marginal likelihood is straightforward for the INAR model. Firstly, efficient data augmentation MCMC algorithms exist for obtaining samples from the posterior distribution in \cite{NSR} (standard INAR model) and \cite{ENSb} (INAR model with explanatory variables). Secondly, the posterior distributions are uni-modal and thus a prior distribution based on a mixture of a Gaussian approximation of the MCMC output and the prior works well. Thirdly, it is straightforward to calculate $\pi (\mathbf{x} | \btheta, \mathbf{x}_{IN})$ since
\begin{eqnarray} \label{eq:inar:2}
P (\mathbf{X} = \mathbf{x} | \btheta, \mathbf{x}_{IN}) &=& \prod_{t=1}^n P (X_t = x_t | \btheta, \mathbf{x}_{-(p-1):(t-1)})  \nonumber \\
&=& \prod_{t=1}^n P (X_t = x_t | \btheta, \mathbf{x}_{(t-p):(t-1)}), \end{eqnarray} where $\mathbf{x}_{a:b} = (x_a, x_{a+1}, \ldots, x_b)$ and
\begin{eqnarray} \label{eq:inar:3} P (X_t = x_t | \btheta, \mathbf{x}_{(t-p):(t-1)}) &=& \sum_{k_1, \ldots, k_p; \sum k_i \leq x_t}
\prod_{j=1}^p \left\{ \binom{x_{t-j}}{k_j} \alpha_j^{k_j} (1-\alpha_j)^{x_{t-j}-k_j} \right\} \nonumber \\ & & \hspace{1cm} \times \frac{\lambda^{x_t - \sum k_i}}{(x_t - \sum k_i)!} \exp (-\lambda). \end{eqnarray}

%
%
%
%
%
%

\subsection{Poisson regression model} \label{ss:TS:pois}

The Poisson regression model (\cite{Zeger}, \cite{DDS}) involves an observed count process $X_1, X_2, \ldots, X_n$ which depends upon a typically unobserved latent process $Y_1, Y_2, \ldots, Y_n$. Specifically in this paper we assume an $AR (p)$ latent process with
\begin{eqnarray} \label{eq:poisreg:1}
X_t | Y_t \sim {\rm Po} (\mu_t \exp (Y_t)) \\
 \label{eq:poisreg:2} Y_t = \sum_{i=1}^p a_i Y_{t-i} + e_t, \end{eqnarray} where $\mu_t = \exp (\mathbf{z}_t^T \bbeta)$ is assumed to depend upon $k$ explanatory variables $\mathbf{z}_t = (1,z_t^1, \ldots, z_t^k)$ and unknown regression coefficients $\bbeta = (\beta_0, \beta_1, \ldots, \beta_k)$ and the $\{e_t\}$'s are independent and identically distributed according to $N(0,1/\tau)$.
The parameters of interest are $\btheta = (\bbeta, \mathbf{a}, \tau)$, where $\mathbf{a} = (a_1, a_2, \ldots,a_p)$. In the absence of explanatory variables $(k=0)$, we will set $\mu_t = \exp (\beta_0) = \phi$ and replace $\bbeta$ by $\phi$.

It is straightforward to construct a data augmentation MCMC algorithm to obtain samples from $\pi (\btheta | \mathbf{x})$, where $\mathbf{x} = (x_1, x_2, \ldots, x_n)$. Details for the case $p=1$ are given below with the extension to $p>1$, where it differs, given in the appendix. Let $\mathbf{y} = (y_1, y_2, \ldots, y_n)$ and $\mathbf{y}_{IN} = (y_{1-p}, y_{2-p}, \ldots, y_0)$. Then we have that
\begin{eqnarray} \label{eq:poisreg:3}
&& \pi (\mathbf{x}, \mathbf{y}, \mathbf{y}_{IN} | \btheta) \nonumber \\ &=& \prod_{t=1}^n \pi (x_t | y_t, \bbeta) \pi (y_t | \mathbf{y}_{(t-p):(t-1)}, \mathbf{a}, \tau) \times \pi (\mathbf{y}_{IN}| \mathbf{a}, \tau) \nonumber \\
&=& \prod_{t=1}^n \prod_{t=1}^n \frac{(\mu_t \exp (y_t))^{x_t}}{x_t!} \exp (- \mu_t \exp (y_t)) \sqrt{\frac{\tau}{2}} \exp \left( - \frac{\tau}{2} \left\{y_t - \sum_{j=1}^p a_j y_{t-j} \right\}^2 \right) \nonumber \\ && \times \pi (\mathbf{y}_{IN}| \mathbf{a}, \tau). \end{eqnarray}
For $p=1$, the distribution of $\mathbf{y}_{IN} =y_0$ is $y_0 \sim N(0, 1/\{ \tau (1-a^2)\})$. Independent priors are chosen for $\bbeta$, $a$ and $\tau$ with $N(\mathbf{m}_\beta, C_\beta)$, $N(m_a, C_a)$ (constrained between $(-1,1)$ to ensure stationarity) and ${\rm Gamma} (A_\tau, B_\tau)$ priors, respectively.  In the absence of explanatory variables a conjugate ${\rm Gamma} (A_\phi, B_\phi)$ prior is used for $\phi$.

The MCMC algorithm is as follows. Update $\bbeta$ as a block using random walk Metropolis as no nice conditional distribution exists. If there are no explanatory variables then $\phi$ can be drawn from its conditional distribution,
\begin{eqnarray} \label{eq:poisreg:5} \phi | \mathbf{a}, \tau, \mathbf{x}, \mathbf{y}, \mathbf{y}_{IN} \sim {\rm Gamma} \left(\sum_{t=1}^n x_t + A_\phi, \sum_{t=1}^n \exp (y_t) + B_\phi \right). \end{eqnarray}
For $\mathbf{a}$ and $\tau$ if we ignore the term $\pi (\mathbf{y}_{IN}| \mathbf{a}, \tau)$, then $\mathbf{a}$ and $\tau$ would have multivariate Normal and Gamma posterior distributions, respectively, given by
\begin{eqnarray} \label{eq:poisreg:6}
\mathbf{a}| \bbeta, \tau, \mathbf{x}, \mathbf{y}, \mathbf{y}_{IN} & \sim & N (\mathbf{M}_a, S_a) \\ \label{eq:poisreg:7}
\tau| \bbeta, \mathbf{a}, \mathbf{x}, \mathbf{y}, \mathbf{y}_{IN} & \sim & {\rm Gamma} \left(A_\tau + \frac{n}{2}, B_\tau + \frac{1}{2} \sum_{t=1}^n \left(y_t - \sum_{j=1}^p a_j y_{t-j} \right)^2 \right), \nonumber \\ \end{eqnarray}
where $\mathbf{Y}$ is a $n \times p$ matrix with $i^{th}$ row $(y_{i-1}, y_{i-2}, \ldots, y_{i-p})$, $S_a = (C_a^{-1} + \tau \mathbf{Y}^T \mathbf{Y})^{-1}$ and $\mathbf{M}_a = S_a (C_a^{-1} \mathbf{m}_a + \tau \mathbf{Y}^T \mathbf{y} )$. Therefore we update $\mathbf{a}$ and $\tau$ by proposing a new value according to \eqref{eq:poisreg:6} and \eqref{eq:poisreg:7}, respectively, accepting the proposed move with probability
\begin{eqnarray} \label{eq:poisreg:8} \min \left\{ 1, \frac{\pi (\mathbf{y}_{IN}| \mathbf{a}^\prime, \tau^\prime)}{\pi (\mathbf{y}_{IN}| \mathbf{a}, \tau)} \right\}. \end{eqnarray}
Finally, the components of $\mathbf{y}$ and $\mathbf{y}_{IN}$ are updated sequentially one at a time using random walk Metropolis. The proposal variance for each $\mathbf{y}$ is tuned automatically using the updating scheme used in \cite{XN14a}, equations (12) and (13).

%
%
%

It is again trivial to obtain a multivariate Gaussian approximation for the posterior distribution, so we turn our attention to estimating $\pi (\mathbf{x})$.  We sample $\btheta =(\bbeta, \mathbf{a},\tau)$ from a mixture of the multivariate Gaussian approximation and the prior but unlike in Section \ref{ss:TS:inar} we can't compute $\pi (\mathbf{x}  | \btheta)$. The solution is to use particle filtering to estimate $\pi (\mathbf{x}  | \btheta)$ as follows. Set $M \geq 1$ and generate $M$ copies of $\mathbf{y}_{IN}$, denoted $\mathbf{s}_0^1, \mathbf{s}_0^2, \ldots, \mathbf{s}_0^M$, from $\pi (\mathbf{y}_{IN} | \btheta)$. For $j=1,2, \ldots, M$, set $w_0^j = 1/M$. Then for $t=1,2,\ldots,n$, we perform the following particle filter steps, for  $j=1,2, \ldots, M$:
\begin{enumerate}
\item Sample $K$ from $\{1,2,\ldots, M\}$ with $P(K=k) = w_{t-1}^k/\sum_{l=1}^M w_{t-1}^l$.
\item Sample $y_t^i \sim N (\sum_{i=1}^p a_i s_{t-1,p+1-i}^k,1/\tau)$.
\item Set $\mathbf{s}_t^j = (s_{t-1,2}^k, \ldots, s_{t-1,p}^k,y_t^j)$ and
\[ w_t^j = \frac{\exp (\mathbf{z}_t^T \bbeta + y_t^j)^{x_t}}{x_t!} \exp ( - \exp (\mathbf{z}_t^T \bbeta + y_t^j)) ( = \pi (x_t | y_t, \bbeta)). \]
\end{enumerate}
Let $P_t = \frac{1}{M} \sum_{j=1}^M w_t^j$. Now provided $\mathbf{s}_{t-1}^1, \ldots, \mathbf{s}_{t-1}^M$ are samples from $\pi (\mathbf{y}_{(t-p):(t-1)} | \mathbf{x}_{1:(t-1)}, \btheta)$, $P_t$ is an unbiased estimate of $P (x_t | \mathbf{x}_{1:(t-1)}, \btheta)$. This is trivially the case for $t=1$ with $P (x_1 | \mathbf{x}_{1:0}, \btheta) = P(x_1 | \btheta)$ and holds for $t>1$ by induction. Therefore $\prod_{t=1}^n P_t$ provides an unbiased estimate of $\pi (\mathbf{x}|\btheta)$.

\subsection{Comparison} \label{ss:TS:comp}

\subsubsection{Introduction}

%
%

We illustrate the methodology with two time series examples. These are the monthly total number of polio cases in the USA from January 1970 to December 1983, \cite{Zeger}, \cite{DDW00a} and the monthly total number of injured logging workers claiming benefit from January 1985 to December 1994, \cite{ZJ}, \cite{ENSb}. In both cases we fit the INAR(1) model, the Poisson regression model with a latent AR(1) process to the data and either an INAR or Poisson regression model with covariates chosen on the basis of the data and initial investigations. Throughout the priors for the INAR(1) model are $U(0,1)$ for $\alpha$ and ${\rm Exp} (1)$ for $\lambda$ and for the priors for the Poisson regression model are ${\rm Exp} (1)$ for $\mu$ and $\tau$ and $N(0,1)$ truncated to $(-1,1)$ for $a$. In all cases the MCMC algorithms were run for 110000 iterations with 10000 iterations discarded as burn-in and the computation of the marginal likelihood is based on 10000 importance sample simulations with $M=1000$. For the importance sampling parameters $\btheta$ are drawn from $q (\btheta) = 0.95 q_N (\btheta) + 0.05 \pi (\btheta)$, where $q_N (\btheta)$ is the probability density function of the multivariate Gaussian distribution with mean vector and covariance matrix given by the MCMC output.

\subsubsection{Polio data}

The polio data has disease case counts ranging from 0 to 14 with the majority being 0 or 1s and a mean of 1.3333. Time series plots of the data are given in \cite{Zeger} and \cite{ENSb}. Fitting the INAR(1) model to the data yielded posterior means (standard deviations) of $0.1877 (0.0469)$ and $1.010 (0.0954)$ for $\alpha$ and $\lambda$, respectively. The estimated log marginal likelihood is $-293.84$. By comparison the Poisson regression model gave an  estimated log marginal likelihood of $-263.33$ with posterior means (standard deviations) of $0.9168 (0.1497)$, $0.5598 (0.1291)$ and $2.031 (0.6087)$ for $\mu$, $a$ and $\tau$, respectively. Therefore there is overwhelming support in favour of the Poisson regression model. Consequently, we fitted a Poisson regression model with covariates to the data to see if a better fit could be achieved. Specifically, we follow \cite{Zeger} and  \cite{DDW00a} in taking
\[ \mathbf{z}_t = \left(1 , \frac{t^\prime}{1000}, \cos \left( \frac{2
\pi t^\prime}{12} \right),  \sin \left( \frac{2 \pi t^\prime}{12}
\right), \cos \left( \frac{2 \pi t^\prime}{6} \right),  \sin \left(
\frac{2 \pi t^\prime}{6} \right) \right), \] where $t^\prime =t
-73$. That is, we assume a linear trend (with intercept
January 1976) and two periods, one of 6 months and the other of 12
months. For $\bbeta = (\beta_1, \ldots, \beta_6)$ we assign independent $N(0,1)$ priors to each $\beta_i$. The estimated log marginal likelihood was $-263.13$ showing only a slight improvement on the Poisson regression model without covariates. The posterior means (standard deviations) of $a$ and $\tau$ were $0.5730 (0.1473)$ and $2.544 (0.8486)$, respectively, similar to those obtained above.
The posterior means and standard deviations of $\bbeta$ are
$(-0.1203, -0.3659,  0.1614,-0.4621, 0.3963,
-0.0037)$ and $(0.1626,0.9253,0.1579,0.1707,0.1401,0.1367)$, respectively. This suggests that most of the covariates do not have a significant effect and supports the findings obtained from comparing the log marginal likelihoods that the simpler model without covariates is adequate.

\subsubsection{Cut injury data}

The cut injury data has counts ranging from 1 to 21 with a mean of 6.1333.  A time series plot of the data is given in \cite{ENSb}. Fitting the INAR(1) model to the data yielded posterior means (standard deviations) of $0.4388 (0.0497)$ and $3.419 (0.3280)$ for $\alpha$ and $\lambda$, respectively. The estimated log marginal likelihood is $-298.3$. By comparison the Poisson regression model gave an  estimated log marginal likelihood of $-306.3$ with  posterior means (standard deviations) of $5.123 (0.7029)$, $0.6892 (0.1017)$ and $7.532 (1.6913)$ for $\mu$, $a$ and $\tau$, respectively. Therefore in this case the preferred model is the INAR(1) model with a Bayes factor of 2984.  The counts are typically higher in the summer months (May to November) than in the winter months (December to April). Consequently, we fitted an INAR(1) model with covariates to the data with $\mathbf{z}_t = (1,s_t)$, where $s_t =1$ for the summer months and $s_t=0$ otherwise. As mentioned in Section \ref{ss:TS:inar}, we take $\alpha^t = \exp (\mathbf{z}_t^T \bbeta) /(1+ \exp (\mathbf{z}_t^T \bbeta))$ and $\lambda^t = \exp ( \mathbf{z}_t^T \bgamma))$ with independent $N(0,1)$ priors for each of the components of $\bbeta$ and $\bgamma$. The estimated log marginal likelihood was $-286.0$ showing a significant improvement on the standard INAR(1) model. The posterior means of $\bbeta$ and $\bgamma$ were $(-0.3361,-0.1230)$ and $(0.8229,0.7027)$, respectively, with corresponding standard deviations $(0.3344, 0.4241)$ and $(0.1871, 0.2116)$, respectively. Thus $\lambda$ varies more between the seasons than $\alpha$, that is, the number of new cases is seasonal whilst the number of injuries carrying from one month to the next is more consistent throughout the year.

\section{Conclusions} \label{S:conc}

In this paper we have introduced a simple three stage algorithm for efficiently estimating the marginal likelihood. The key components are an MCMC algorithm for obtaining samples from the posterior distribution, $\pi (\btheta | \mathbf{x})$, an approximating distribution $q (\btheta )$ to sample from and an effective estimate of the likelihood $\pi (\mathbf{x}| \btheta)$. The first observation is whilst an MCMC algorithm will often be relatively straightforward to construct, alternative methods for sampling from the posterior distribution could be equally considered. Moreover, it is not important if a sample from an approximate posterior distribution is used since all that is required for computation of the marginal likelihood is to be able to make a reasonable choice of $q(\cdot)$. The key limitation to using this approach is effective estimation of the likelihood $\pi (\mathbf{x}| \btheta)$ in cases where it is not analytically tractable. For the examples in this paper we have been able to exploit the temporal nature of the data to use filtering methods to estimate $\pi (\mathbf{x}| \btheta)$ which will be applicable more generally to longitudinal and time series data. Furthermore the importance sampling and the associated estimation of the likelihood is trivially parallelisable which can be utilised to speed up implementation.  In cases where the likelihood can easily be computed the algorithm becomes a simple add-on to MCMC to compute the marginal likelihood.

\section*{Acknowledgements}

PT was supported by a University of Warwick PhD scholarship. NA was supported by a PhD scholarship from the Saudi Arabian Government.
PN, SS and TM would like to thank the organisers of the Design and Analysis of Infectious Disease Studies workshop at Oberwolfach (November 2013), where many helpful discussions took place.

\appendix
\section{Epidemic model}

\subsection{Marginal likelihood estimation}
\label{sup:S:mlcalculation}

In this section we briefly overview  alternative techniques for estimating the likelihood. However first we give the full list of proposals distributions based on a fitted multivariate normal distribution with mean $\bmu$ and covariance matrix $\bSigma$ used in the importance sampling.
These are
\begin{enumerate}[{1)}] \itemsep=0pt\parskip=-8pt
\item $\mathrm{IS}_{N_1}$ : $q(\btheta)=N(\btheta; \bmu,\bSigma)$,\\
\item $\mathrm{IS}_{N_2}$ : $q(\btheta)=N(\btheta; \bmu,2\bSigma)$,\\
\item $\mathrm{IS}_{N_3}$ : $q(\btheta)=N(\btheta; \bmu,3\bSigma)$,\\
\item $\mathrm{IS}_{N_4}$ : $q(\btheta)=N(\btheta; \bmu,4\bSigma)$,\\
\item $\mathrm{IS}_{\mathrm{mix}}$ :  $q(\btheta)=0.95\times N(\btheta; \bmu,\bSigma)+0.05\times \pi(\btheta)$,\\
\item $\mathrm{IS}_{t_4}$ :  $q(\btheta)=t_4(\btheta; \bmu,\bSigma)$,\\
\item $\mathrm{IS}_{t_6}$ :  $q(\btheta)=t_6(\btheta; \bmu,\bSigma)$,\\
\item $\mathrm{IS}_{t_8}$ :   $q(\btheta)=t_8(\btheta; \bmu,\bSigma)$,\\
\item $\mathrm{IS}_{t_{10} }$ :   $q(\btheta)=t_{10}(\btheta; \bmu,\bSigma)$,
\end{enumerate}
where $t_d(\btheta,\bmu,\bSigma)$ is the density of the multivariate Student's $t$ distribution with $d$ degrees of freedom, mean $\bmu$ and covariance matrix $\frac{d}{d-2}\bSigma$ (if $d>2$).

\subsubsection{Marginal likelihood estimation via the harmonic mean}

The harmonic mean (HM) estimator \cite{Newton} can be computed directly from MCMC output, which has lead to its widespread use. When data augmentation is used, the parameter vector comprises latent variable $\mathbf{y}$ as well as the model parameters $\btheta$. The marginal likelihood $\pi(\mathbf{x})$ can be approximated by the sample harmonic mean of the likelihoods,
\begin{eqnarray}
\hat{P}_{HM} (\mathbf{x}) =  \Bigg[ \frac{1}{N} \sum_{i=1}^N \frac{1}{P (\mathbf{x} | \mathbf{y}_i, \btheta_i)}\Bigg]^{-1}
\end{eqnarray}
based on $N$ draws $(\mathbf{y}_1, \btheta_1), (\mathbf{y}_2, \btheta_2), \ldots, (\mathbf{y}_N , \btheta_N)$ from the joint posterior $\pi(\mathbf{y}, \btheta | \mathbf{x})$. Although asymptotically consistent, the harmonic mean estimator is known to exhibit large or even infinite variance for some models.

\subsubsection{Marginal likelihood estimation via Chib's method}

Chib's method for estimating the marginal likelihood \cite{Chib,CJ01} is based on the observation that,
$$\pi(\mathbf{x})=\frac{\pi(\mathbf{x}|\btheta)\pi(\btheta)}{\pi(\btheta|\mathbf{x})},$$
for all $\btheta$ in the support of the posterior. For fixed $\btheta=\btheta^*$ the log marginal likelihood can be estimated by
\begin{eqnarray}
\log \widehat{P}_{\text{Chib}}(\mathbf{x}) =  \log \pi (\mathbf{x} | \btheta^*) + \log \pi(\btheta^*) - \log \widehat{\pi} (\btheta^* | \mathbf{x})
\end{eqnarray}
The prior and likelihood can be easily evaluated at $\btheta^*$. The posterior density is estimated by breaking the parameter vector into appropriate blocks. For blocks that can be updated using a Gibb's step the normalising constant of the full conditional distribution is known and so MCMC samples of the remaining blocks can be used to estimate the normalising constant of the posterior. For components of $\btheta$ that are updated using Metropolis-Hastings, the required normalising constant can be estimated from the acceptance probabilities of a jump from the current state to $\btheta^*$.

For the model described in Section 3.2 we decompose the parameter vector into $(\mathbf{y},\btheta_1,\btheta_2,\btheta_3)$, where $\btheta_1 = (k_1, k_2, \beta_{11}, \beta_{12}, \beta_{21}, \beta_{22}, \mu_1, \mu_2, w)$, $\btheta_2 = \pi_1$ and $\btheta_3 = \pi_2$. The posterior density is then factorised as
\begin{eqnarray*}
\pi(\mathbf{y}^*, \btheta^* | \mathbf{x}) = \pi(\mathbf{y}^*|\mathbf{x})\,\pi(\btheta_1^*|\mathbf{x}, \mathbf{y}^*)\,\pi(\btheta_2^*|\mathbf{x}, \mathbf{y}^*, \btheta_1^*)\,\pi(\btheta_3^*|\mathbf{x}, \mathbf{y}^*, \btheta_1^*, \btheta_2^*).
\end{eqnarray*}
To calculate each term in this product, a separate MCMC chain is run in which only the unconditioned blocks of $(\mathbf{y},\btheta)$ are updated and the remaining blocks are fixed at $\btheta^*$.

\subsubsection{Marginal likelihood estimation via power posteriors}

The Power Posterior (PP) approach to estimating the marginal likelihood \cite{Friel} uses samples from the power posterior, defined as
\begin{eqnarray*}
\pi_t(\btheta | \mathbf{x}) \propto \pi(\mathbf{x} | \btheta)^t \, \pi(\btheta)
\end{eqnarray*}
 where $t \in [0,1]$ is a temperature parameter. Borrowing ideas from path sampling allows the log of the marginal likelihood to be represented in terms of the thermodynamic integral
 \begin{eqnarray*}
\log \pi(\mathbf{x}) = \int_0^1 \mathrm{E}_{\btheta | \mathbf{x}, t} \{ \log  \pi(\mathbf{x} | \btheta) \} \, dt,
\end{eqnarray*}
 where the expectation of the mean deviance is taken with respect to the power posterior at temperature $t$, where $t$ moves from 0 to 1. The integral can be calculated numerically by discretising the temperature range as $0= t_0 < t_1 < \ldots < t_n = 1$, and then the log marginal likelihood can be approximated by the trapezium rule,
 \begin{eqnarray*}
\log \widehat{P}_{PP}(\mathbf{x}) = \sum_{i=0}^{n-1}(t_{i+1} - t_i) \frac{\mathrm{E}_{\btheta | \mathbf{x}, t_{i+1}} \{ \log  \pi(\mathbf{x} | \btheta) \}  + \mathrm{E}_{\btheta | \mathbf{x}, t_i} \{ \log  \pi(\mathbf{x} | \btheta) \}} {2}.
\end{eqnarray*}
For each $t_i$, samples from the power posterior $\pi_{t_i}(\btheta | \mathbf{x})$ can be used to obtain an estimate of the required expectation
 $\mathrm{E}_{\btheta | \mathbf{x}, t_i} \{ \log  \pi(\mathbf{x} | \btheta) \}$. Metropolis within Gibbs sampling was used to obtain samples from the power posterior at each temperature $t>0$.

The variability of the power posterior estimator depends on the chosen number and spacing of the $t_i$'s. Choosing a large number of temperatures, the estimation of the log marginal likelihood requires considerably more computational effort. Moreover, the precision of the estimate is sensitive to the number of samples used and the mixing of the MCMC sampler.

\subsection{Simulation study}

\subsubsection{Temperatures for the power posteriors}

\begin{figure}[h!]
\centering
\footnotesize
\begin{tikzpicture}[x=1pt,y=1pt]
\definecolor{fillColor}{RGB}{255,255,255}
\path[use as bounding box,fill=fillColor,fill opacity=0.00] (0,0) rectangle (289.08,216.81);
\begin{scope}
\path[clip] ( 33.60, 33.60) rectangle (277.08,204.81);
\definecolor{drawColor}{RGB}{0,0,0}

\path[draw=drawColor,line width= 0.4pt,line join=round,line cap=round] ( 42.62, 39.94) --
	( 46.75,155.23) --
	( 52.12,160.72) --
	( 58.07,162.33) --
	( 64.44,163.08) --
	( 71.14,163.56) --
	( 78.12,163.85) --
	( 85.33,164.05) --
	( 98.98,164.30) --
	(144.07,164.63) --
	(181.59,164.79) --
	(192.98,170.55) --
	(204.10,173.80) --
	(214.90,175.89) --
	(225.35,182.19) --
	(235.38,187.05) --
	(244.92,190.26) --
	(253.84,193.06) --
	(261.87,196.34) --
	(268.06,198.47);
\end{scope}
\begin{scope}
\path[clip] (  0.00,  0.00) rectangle (289.08,216.81);
\definecolor{drawColor}{RGB}{0,0,0}

\path[draw=drawColor,line width= 0.4pt,line join=round,line cap=round] ( 33.60, 33.60) --
	(277.08, 33.60) --
	(277.08,204.81) --
	( 33.60,204.81) --
	( 33.60, 33.60);
\end{scope}
\begin{scope}
\path[clip] (  0.00,  0.00) rectangle (289.08,216.81);
\definecolor{drawColor}{RGB}{0,0,0}

\node[text=drawColor,rotate= 90.00,anchor=base,inner sep=0pt, outer sep=0pt, scale=  1.00] at ( -4.80,119.21) { };
\end{scope}
\begin{scope}
\path[clip] (  0.00,  0.00) rectangle (289.08,216.81);
\definecolor{drawColor}{RGB}{0,0,0}

\path[draw=drawColor,line width= 0.4pt,line join=round,line cap=round] ( 42.62, 33.60) -- (268.06, 33.60);

\path[draw=drawColor,line width= 0.4pt,line join=round,line cap=round] ( 42.62, 33.60) -- ( 42.62, 30.18);

\path[draw=drawColor,line width= 0.4pt,line join=round,line cap=round] ( 87.71, 33.60) -- ( 87.71, 30.18);

\path[draw=drawColor,line width= 0.4pt,line join=round,line cap=round] (132.80, 33.60) -- (132.80, 30.18);

\path[draw=drawColor,line width= 0.4pt,line join=round,line cap=round] (177.88, 33.60) -- (177.88, 30.18);

\path[draw=drawColor,line width= 0.4pt,line join=round,line cap=round] (222.97, 33.60) -- (222.97, 30.18);

\path[draw=drawColor,line width= 0.4pt,line join=round,line cap=round] (268.06, 33.60) -- (268.06, 30.18);

\node[text=drawColor,anchor=base,inner sep=0pt, outer sep=0pt, scale=  1.00] at ( 42.62, 18.00) {0};

\node[text=drawColor,anchor=base,inner sep=0pt, outer sep=0pt, scale=  1.00] at ( 87.71, 18.00) {0.2};

\node[text=drawColor,anchor=base,inner sep=0pt, outer sep=0pt, scale=  1.00] at (132.80, 18.00) {0.4};

\node[text=drawColor,anchor=base,inner sep=0pt, outer sep=0pt, scale=  1.00] at (177.88, 18.00) {0.6};

\node[text=drawColor,anchor=base,inner sep=0pt, outer sep=0pt, scale=  1.00] at (222.97, 18.00) {0.8};

\node[text=drawColor,anchor=base,inner sep=0pt, outer sep=0pt, scale=  1.00] at (268.06, 18.00) {1};

\node[text=drawColor,anchor=base,inner sep=0pt, outer sep=0pt, scale=  1.00] at (155.34,  2.40) {Temperature};

\path[draw=drawColor,line width= 0.4pt,line join=round,line cap=round] ( 33.60, 48.22) -- ( 33.60,185.69);

\path[draw=drawColor,line width= 0.4pt,line join=round,line cap=round] ( 33.60, 48.22) -- ( 30.18, 48.22);

\path[draw=drawColor,line width= 0.4pt,line join=round,line cap=round] ( 33.60, 82.59) -- ( 30.18, 82.59);

\path[draw=drawColor,line width= 0.4pt,line join=round,line cap=round] ( 33.60,116.95) -- ( 30.18,116.95);

\path[draw=drawColor,line width= 0.4pt,line join=round,line cap=round] ( 33.60,151.32) -- ( 30.18,151.32);

\path[draw=drawColor,line width= 0.4pt,line join=round,line cap=round] ( 33.60,185.69) -- ( 30.18,185.69);

\node[text=drawColor,rotate= 90.00,anchor=base,inner sep=0pt, outer sep=0pt, scale=  1.00] at ( 25.20, 48.22) {-3000};

\node[text=drawColor,rotate= 90.00,anchor=base,inner sep=0pt, outer sep=0pt, scale=  1.00] at ( 25.20, 82.59) {-2500};

\node[text=drawColor,rotate= 90.00,anchor=base,inner sep=0pt, outer sep=0pt, scale=  1.00] at ( 25.20,116.95) {-2000};

\node[text=drawColor,rotate= 90.00,anchor=base,inner sep=0pt, outer sep=0pt, scale=  1.00] at ( 25.20,151.32) {-1500};

\node[text=drawColor,rotate= 90.00,anchor=base,inner sep=0pt, outer sep=0pt, scale=  1.00] at ( 25.20,185.69) {-1000};

\node[text=drawColor,rotate= 90.00,anchor=base,inner sep=0pt, outer sep=0pt, scale=  1.00] at (  9.60,119.21) {Expected deviance};
\end{scope}
\end{tikzpicture}
   \caption{Expected deviance against temperature for model $\mathcal{M}_1$ estimated using power posteriors.}
    \label{fig:ExpDevia}
 \end{figure}
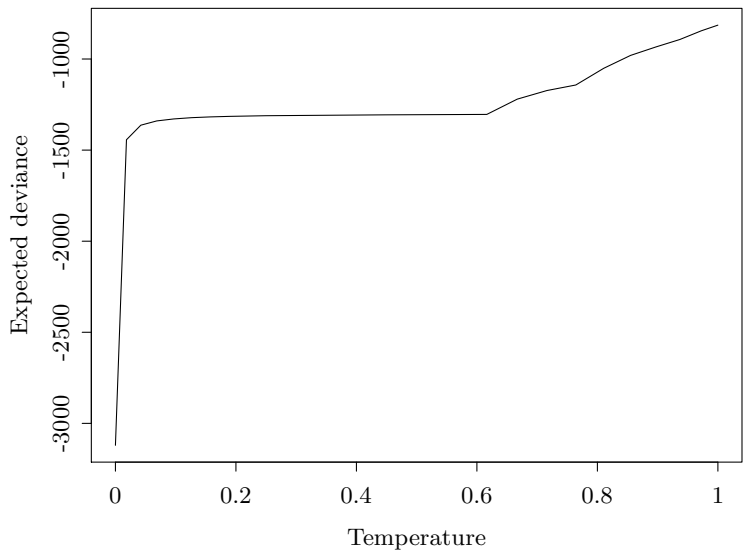

In Friel and Pettitt \cite{Friel} chose a geometric spacing of the temperatures, $t_l= (l/n)^c$, for $l = 0, 1, \ldots, n$, with $c>1$, which places many of the temperatures close to zero. This scheme is preferable in cases where the expected deviance has a sharp increase near zero before leveling off. However, in our case, the curve of the expected deviance is not convex (Figure \ref{fig:ExpDevia}). After some pilot analysis (not counted in the computation cost) we chose to use 20 partitions of the unit line, placing more temperatures around zero and the other sharp change.

\subsubsection{Monte Carlo standard errors of log marginal likelihood estimates}

Table \ref{tab:Comparisonsds} gives the Monte Carlo standard errors of log marginal likelihood estimates for different number of Markov chain samples. The results show that increasing the number of MCMC samples $n$, led to a decrease in the Monte Carlo standard errors of order $\mathcal{O}(\sqrt n)$, see Table 1 indicating that the variance of the corresponding estimators is finite.

\begin{table}[h!]
\centering
\begin{tabular}{c |c  c| cc | cc }
  \hline \hline
 \multirow{2}{*}{Method} & MC & MC & MC & MC & MC & MC
 \\
 & samples & error  & samples & error  &  samples& error\\ \hline
 $\mathrm{IS}_{N_1}$& 10000& 0.053 & 25000 & 0.033 & 50000& 0.025 \\ [0.5ex]
 $\mathrm{IS}_{N_2}$& 10000 & 0.064  & 25000& 0.036& 50000 & 0.025 \\ [0.5ex]
 $\mathrm{IS}_{N_3}$& 10000 & 0.107  & 25000& 0.061 & 50000& 0.042 \\ [0.5ex]
 $\mathrm{IS}_{N_4}$& 10000 & 0.174  & 25000& 0.143 & 50000& 0.106 \\ [0.5ex]
 $\mathrm{IS}_{\mathrm{mix}}$ & 10000& \textbf{0.030}  & 25000& \textbf{0.018} & 50000& \textbf{0.012} \\ [0.5ex]
   $\mathrm{IS}_{t_4}$ & 10000& 0.037  & 25000& 0.025 & 50000& 0.020 \\ [0.5ex]
  $\mathrm{IS}_{t_6}$& 10000 & 0.064  & 25000& 0.034 & 50000& 0.023 \\ [0.5ex]
  $\mathrm{IS}_{t_8}$ & 10000& 0.034  & 25000& 0.035 & 50000& 0.024 \\ [0.5ex]
  $\mathrm{IS}_{t_{10}}$ & 10000& 0.041  & 25000& 0.061& 50000 & 0.045 \\ [0.5ex]
  Chib & 8000& 0.736 & 20000 & 0.486 & 40000 & 0.312 \\ [0.5ex]
  PP & 20$\times$1600&  2.906 & 20$\times$2150& 1.936 & 20$\times$3200& 1.547 \\ [0.5ex]
  HM & 37000& 5.548 & 50000& 5.331 &72000& 4.850 \\
\hline \hline
\end{tabular}
\caption{Monte Carlo standard errors of log marginal likelihood estimates for different number of Markov chain samples. Standard errors are given across 50 replicates for each of the methods. Each of the methods have roughly the same computational cost.}
\label{tab:Comparisonsds}
\end{table}

\subsection{Model comparison} \label{ss:S:comp}

In this section we provide further details and examples of the strength of the importance sampling technique in answering model comparison questions. We begin by providing details of the reversible jump algorithm for heterogeneity in community acquisition rates. This is followed by investigations of model comparison for heterogeneity in household transmission rates, density-dependence in within-household transmission and the amount of missing data.

\subsubsection{Reversible jump MCMC}

Reversible jump MCMC \cite{Green} provides a framework for constructing MCMC algorithms that can jump between states with different dimensions. This allows the model indicator to be treated as a parameter to be estimated from the data like any other. Although the Bayes factors can be calculated from the posterior probabilities in favour of each model, the marginal likelihoods themselves cannot be obtained. The main difficulty with RJMCMC lies in designing efficient proposals to jump between models and their associated parameters.

Here we wish to compare model $\mathcal{M}_1$ described in Section 3.2 with the nested model $\mathcal{M}_2$, in which the community acquisition rates for adults and children are equal, i.e. $k_1=k_2=k$ for some $k$ (Section 3.6). When in model $\mathcal{M}_1$, we propose a move to $\mathcal{M}_2$ with probability 0.5, in which the joint community acquisition rate $k$ is set to $k = \frac{L_1 \, k_1 + L_2 \, k_2}{L_1+L_2}$, where $L_1$ is the total number of children and $L_2$ is the total number of adults. The Jacobian of the transformation is $\frac{L_1 \, L_2}{L_1+L_2}$. For the reverse move, we need to increase the dimension of the parameter vector, therefore an auxiliary random variable $U$ is required. Let $U \sim N(0, \sigma^2)$ with $\sigma^2$ fixed but well chosen. Then we set $k_1 =  k  + \frac{u}{L_1}$ and $k_2 =  k  - \frac{u}{L_2}$. The Jacobian of the transformation is then $\frac{L_1+L_2}{L_1 \, L_2}$. The  acceptance probability of jumping from $\mathcal{M}_1$ to $\mathcal{M}_2$, is given by $\min(1, A_{12})$ where
 \begin{eqnarray*}
  A_{12} = \frac{ \pi(\mathcal{M}_2, \boldsymbol{\phi}_2 | \mathbf{x}) \, \pi(\mathcal{M}_2)}{ \pi(\mathcal{M}_1, \boldsymbol{\phi_1} | \mathbf{x}) \, \pi(\mathcal{M}_1)} \Bigg (\frac{1}{\sigma \sqrt{2\pi}} e^{-\frac{1}{2\sigma^2}{\left( \frac{L_1 L_2 (k_1 - k_2)}{L_1+L_2}\right)^2}}  \Bigg) \frac{L_1 \, L_2}{L_1+L_2},
\end{eqnarray*}
where $\boldsymbol{\phi}_1 = (k_1, k_2, \beta_{11}, \beta_{12}, \beta_{21},\beta_{22}, w, \mu_1, \mu_2,\pi_1, \pi_2, \mathbf{y})$ and $  \boldsymbol{\phi}_2 = (k,\beta_{11},\allowbreak \beta_{12}, \beta_{21}, \beta_{22}, w, \mu_1, \mu_2, \pi_1, \pi_2, \mathbf{y})$. For the reciprocal move from model $\mathcal{M}_2$ to $\mathcal{M}_1$, the probability of accepting the jump is given by $\min(1, A_{21})$ where
 \begin{eqnarray*}
  A_{21} = \frac{ \pi(\mathcal{M}_1, \boldsymbol{\phi}_1 | \mathbf{x}) \, \pi(\mathcal{M}_1)}{\pi(\mathcal{M}_2, \boldsymbol{\phi_2} | \mathbf{x}) \, \pi(\mathcal{M}_2)} \Bigg( \frac{1}{\sigma \sqrt{2\pi}} e^{ -\frac{u^2}{2\sigma^2}} \Bigg ) ^{-1} \frac{L_1+L_2}{L_1 \, L_2}.
\end{eqnarray*}
In addition to the model-switching step, the within-model parameters are updated using a standard MCMC algorithm that employs both Gibbs sampler updates and random walk Metropolis steps with a Gaussian proposal density centred to the current value.

\subsubsection{Heterogeneity in household transmission rates}

We wish to evaluate whether or not there is heterogeneity in the household transmission rates. More precisely, we wish to compare the full model $\mathcal{M}_1$ with the special case in which the within-household acquisition rates are identical between the two age groups, i.e. $\beta_{11} = \beta_{12} = \beta_{21} = \beta_{22} = \beta$ (say), which we call model $\mathcal{M}_3$. This kind of question is extremely challenging to answer using reversible jump methodology because it is difficult to move efficiently between models when this involves a large change in dimension. Again we generated two datasets, one from $\mathcal{M}_1$ using the parameters given in Section 3.5, and one from $\mathcal{M}_3$ with $\beta=0.0515$, the average of $\beta_{11}, \beta_{12}, \beta_{21}$ and $\beta_{22}$. For both datasets, we calculated Bayes factors using importance sampling, Chib's method, power posteriors and the harmonic mean. Our objective was to check that the correct model was chosen by the Bayes factors criterion in this setting.

Table \ref{tab:Comparisonbs} presents the marginal likelihood estimates and the corresponding Monte Carlo standard errors for each method, where bold entries show the preferred model. The importance sampling, Chib and power posterior methods all agreed and were able to discriminate the true model. The estimates of the log marginal likelihoods are similar within Monte Carlo error, with importance sampling being the most precise.  As was previously observed in Section 3.5, the harmonic mean overestimated the log marginal likelihoods and yielded inaccurate results, favouring the wrong model in both scenarios.

\begin{table}[h!]
\centering
\begin{tabular}{ c |c |r | r| r}
  \hline \hline
  Simulation &  \multirow{2}{*}{Method}  & \multicolumn{1}{c|}{Log marginal of} & \multicolumn{1}{c|}{Log marginal of} &  \multicolumn{1}{c}{\multirow{2}{*}{  log$B_{MR}$} } \\
 designs & & \multicolumn{1}{c|}{reduced model} &\multicolumn{1}{c|}{ main model} &
  \\
  \hline
& $\mathrm{IS}_{\mathrm{mix}}$ & \textbf{-1267.102} (0.018) & -1268.843 (0.020) & 1.742 (0.031) \\  [0.5ex]
Reduced &Chib & \textbf{-1266.999} (0.261) & -1268.075 (0.619) & 1.190 (0.729) \\  [0.5ex]
 Data &PP & \textbf{-1262.957} (1.926) & -1266.150 (2.107) & 3.215 (2.465) \\  [0.5ex]
 & HM & -931.320 (3.882) & \textbf{-929.168} (5.444) & -3.562 (6.507) \\  [0.5ex] \hline
  & $\mathrm{IS}_{\mathrm{mix}}$ & -1512.107 (0.011) & \textbf{-1505.058} (0.015) & -7.048 (0.019) \\  [0.5ex]
Main & Chib & -1512.110 (0.326) & \textbf{-1505.021} (0.290)  & -7.156 (0.445) \\  [0.5ex]
 Data  &PP & -1509.138 (2.003) & \textbf{-1500.616} (2.089) & -8.833 (2.495) \\  [0.5ex]
 & HM & \textbf{-1184.755} (5.150) & -1195.668 (6.252) & 9.273 (7.552) \\  [0.5ex]
\hline \hline
\end{tabular}
\caption{Bayes factors and log marginal likelihoods of the main and reduced models for the two simulation designs. The Monte Carlo standard errors over 50 replicates are shown in parentheses.}
\label{tab:Comparisonbs}
\end{table}

\subsubsection{Density-dependence in within-household transmission}

Melegaro \emph{et al.} \cite{Melegaro} investigated the relationship between transmission rates and household size via the density correction factor $(z-1)^{w}$ in the transmission rates, (7) in the main text, where $z$ is the household size. Since their confidence interval for $w$ included 1 they were unable to determine whether transmission increased ($w<1$) or decreased ($w>1$) with household size. Moreover, the value $w=1$ corresponds to frequency dependent transmission, where the average number of contacts is the same irrespective of household size. We wish to determine whether frequency dependent transmission ($w=1$, which we call model $\mathcal{M}_4$) could be identified from the data.

\begin{figure}[h!]
\centering
\input{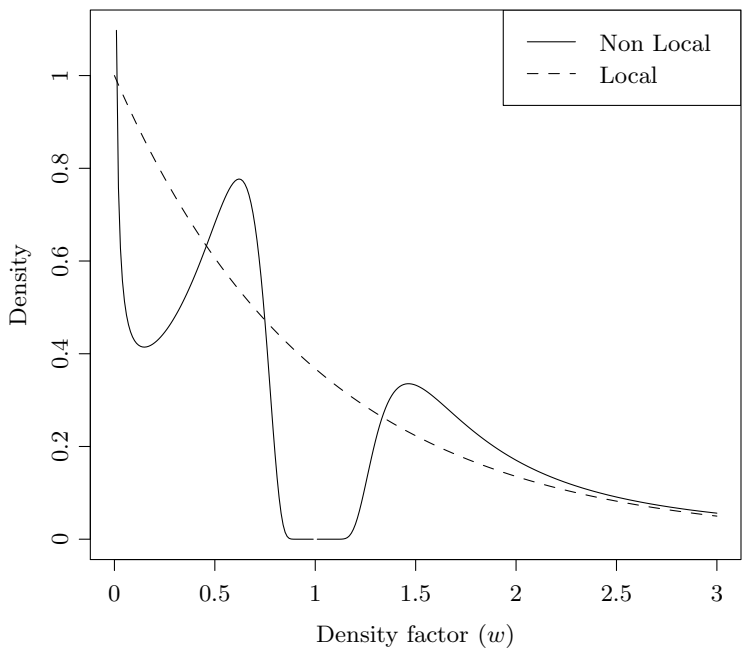}
    \caption{Prior densities on the density correction factor $w$.}
 \label{fig:Priorsw}
 \end{figure}

Bayesian model comparison problems of this kind often suffer from Lindley's paradox, where the choice of prior for $w$ in the more complex model has undue influence on the resulting Bayes factor. To reduce (but not remove) the impact of Lindley's paradox we consider two priors for $w$ in $\mathcal{M}_1$: Gamma(1,1) (referred as the local prior) and  the inverse moment prior for $\log w$ (referred as the non local prior), with densities respectively given by
\begin{align*}
\pi_{L}(w) &= \frac{b^{a} w^{a-1} e^{-w b}}{\Gamma(a)}, \quad a=1, b=1;
\\
\pi_{NL}(w) &= \frac{\kappa \tau^{\nu/2}}{w \, \Gamma(\nu /2\kappa)} \big( \log (w) \big)^{-(\nu +1)} \exp \Bigg[- \Bigg \{ \frac{ \big(\log(w) \big)^2 }{\tau} \Bigg \}^{-\kappa} \Bigg],
\end{align*}
with $\kappa = 1,\, \nu=1$ and $\tau=0.173$ (for more details see \cite{Rossell}). The density functions of the two priors are shown in Figure \ref{fig:Priorsw}. The figure illustrates the fact that the non local prior has density zero at $w=1$.

\begin{figure}[h!]
\centering
    \input{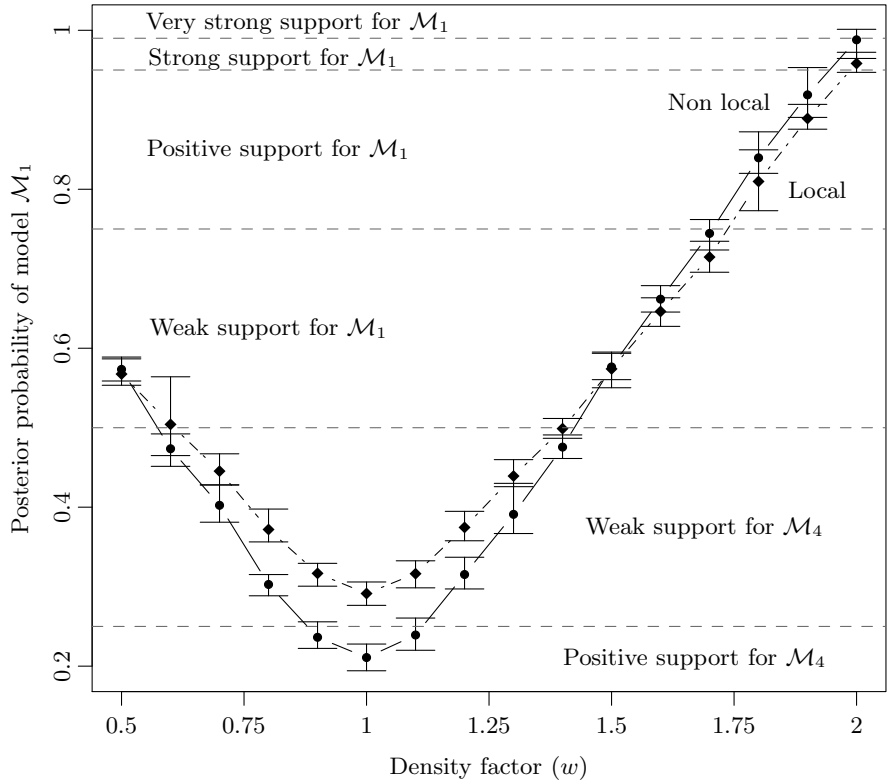}
    \caption{Posterior probability of the full model using two different prior specifications; the local prior ( \protect\rule[0.65ex]{0.15cm}{.4pt} $\cdot$ \protect\rule[0.65ex]{0.15cm}{.4pt} $\cdot$ \protect\rule[0.65ex]{0.15cm}{.4pt}) and the non local prior (\protect\rule[0.7ex]{1.2cm}{.4pt}). Error bars represent the Monte Carlo standard error based on 100 simulations.}
 \label{fig:Bayesw}
 \end{figure}

To determine if evidence in favour or against $\mathcal{M}_4$ could be determined from the study of Melegaro \emph{et al.} we simulated datasets of equivalent size with values for $w$ from 0.5 through to 2, increasing by 0.1 each time. For each value of $w$ we obtained an estimate of the posterior probability of  $\mathcal{M}_1$ along with its standard error, based on 100 simulated datasets. Results are shown in Figure \ref{fig:Bayesw}. For values of $w$ close to 1, the non local prior provided on average stronger evidence in favour of the simple model even though model $\mathcal{M}_1$ was technically the correct model. For values of $w$ within the interval [0.6, 1.4] both priors supported $\mathcal{M}_4$, but only the non local prior provided positive support for $\mathcal{M}_4$. Whereas when $w$ went from 1.5 to 2, both priors favoured $\mathcal{M}_1$, with the non local prior providing equal or higher posterior probability in favour of the correct model than the alternative local prior. Melegaro \emph{et al.} \cite{Melegaro} estimated $w=1.18$ and in this region we expect weak support for frequency dependent transmission, model $\mathcal{M}_4$.

\subsubsection{Amount of missing data}

In this section we wish to assess the accuracy and efficiency of the proposed method as a function of the total number of hidden states. One way to vary the amount of missing data without diluting the information content of the dataset is to vary the time interval $\delta t$. The larger $\delta t$ is, the smaller the number of hidden states that need to be imputed. For example, when $\delta t = 1$ 60840 hidden states need to be imputed, whereas we have only 3900 when $\delta t=10$. For $\delta t= 1, 2, \ldots, 10$, we generated 10 synthetic datasets according to $\mathcal{M}_1$. For each dataset we fitted 10 different models, one for each possible value of $\delta t$, and calculated the log marginal likelihood.

 \begin{figure}[h!]
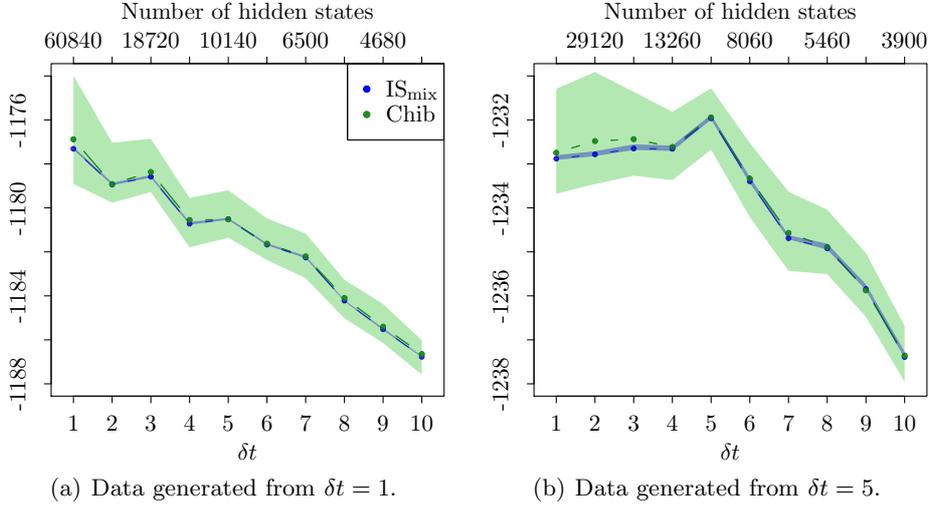
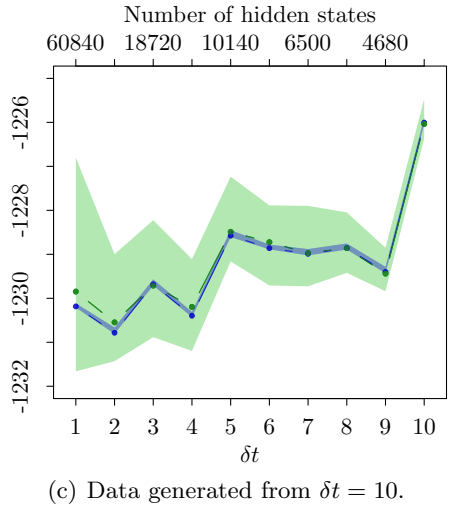

\centering
    \subfigure[Data generated from $\delta t = 1$.]{\input{Timeint1.tex}}
    \subfigure[Data generated from $\delta t = 5$.]{\input{Timeint5.tex}} \newline \newline
     \subfigure[Data generated from $\delta t = 10$.]{\input{Timeint10.tex}}
    \caption{Sensitivity to $\delta t$, the time interval. Log marginal likelihood estimation and its corresponding 95$\%$ credible interval using data generated by (a) $\delta t = 1$, (b) $\delta t=5$ and (c) $\delta t = 10$.}
 \label{fig:Timeintdt}
 \end{figure}

For brevity, Figure \ref{fig:Timeintdt} presents results only for the data generated by $\delta t = 1, 5, 10$. In all three cases, the log marginal likelihood curves are peaked at the true value of $\delta t$, the one used to create the data (Figure \ref{fig:Timeintdt}). The marginal likelihood estimates from Chib's method are also maximized at true values, but since the standard errors are much higher, more samples would be required to distinguish between the competing models.

 \begin{figure}[h!]
\centering
        \input{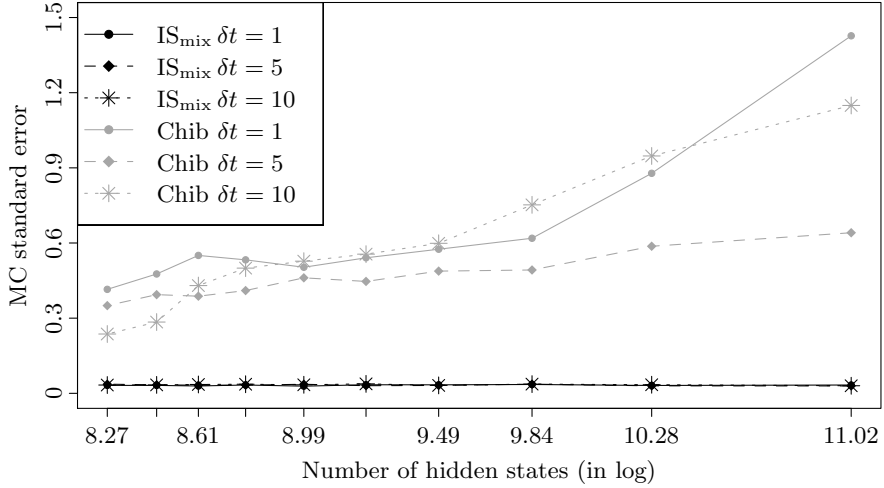}
          \caption{Monte Carlo standard error of the proposed importance sampling   and Chib's methods for different number of hidden states and values of time interval  $\delta t$. }
             \label{fig:Timeintsds}
\end{figure}

Figure \ref{fig:Timeintsds} shows how the Monte Carlo standard errors in Chib's method and the importance sampling method increase as a function of the total number of hidden states. The graph shows that the Monte Carlo standard errors from the importance sampling method appear very stable as the dimensionality of the hidden states is increase.

\section{Time Series models}


\subsection{Poisson regression model: $p >1$} \label{ss:TS:pois}

The main complication with extending the MCMC given in the main text from a latent $AR(1)$ process to a latent $AR(p)$ process is that the distribution of $\mathbf{y}_{IN}$, the stationary distribution of $p$ consecutive observations from the latent process, is non-trivial. (Remember that for $p=1$, $\Sigma = 1/\{ \tau (1-a^2)\}$ and $\mathbf{y}_{IN} =y_0 \sim N(0, 1/\{ \tau (1-a^2)\})$. Let $A$ denote the $p \times p$ square matrix with first row equal to $\mathbf{p}$, $A_{(i+1),i} =1$ $(i=1,\ldots,p-1)$ and all other entries equal to 0. Then, letting $S$ denote the  $p \times p$ square matrix  with all entries equal to 0 except $S_{1,1} =1/\tau$, we have that
\begin{eqnarray} \label{eq:poisreg:4}
\mathbf{y}_{IN} &=& A \mathbf{y}_{-p:-1} + N (\mathbf{0}, S) \nonumber \\
&=& N(0, \Sigma), \mbox{ say}, \end{eqnarray} where $\Sigma= \sum_{j=0}^\infty A^j S (A^j)^T$. A natural prior for $\mathbf{a}$ is $N(\mathbf{m}_a, C_a)$, constrained, if desirable, to ensure that the latent process is stationary.

\end{document}